# Robust Image Watermarking in the Wavelet Domain for Copyright Protection

Hamed Dehghan, S. Ebrahim Safavi
Department of Electrical Engineering, Sharif University of Technology, Tehran, Iran.
Email : {dehghan, safavi}@ee.sharif.edu

*Abstract:* In this paper a new approach to image watermarking in wavelet domain is presented. The idea is to hide the watermark data in blocks of the block segmented image. Two schemes are presented based on this idea by embedding the watermark data in the low pass wavelet coefficients of each block. Due to low computational complexity of the proposed approach, this algorithm can be implemented in real time. Experimental results demonstrate the imperceptibility of the proposed method and its high robustness against various attacks such as filtering, JPEG compression, cropping, noise addition and geometric distortions.

## 1. INTRODUCTION

Due to the increasing use and ease of manipulation of digital media such as audio, image, and video over the internet, the problem of ownership protection has become increasingly important. Embedding a signature, or watermark, into the multimedia signal, is a solution which has been developed in recent years. Digital watermarking allows owners or providers to hide an invisible and robust message inside multimedia content, which is able to be retrieved later [1].

A watermarking scheme should balance between three requirements: robustness, capacity and imperceptibility of embedded data [2]. But, the main challenge in ownership verification system for copyright protection is the robustness of the watermarking techniques against various attacks [3]. The attacks that are usually encountered in image watermarking methods are filtering, JPEG compression, cropping, noise addition and geometric distortions such as rotation and scaling. Although for the first two types of attacks many robust methods were proposed [4-14], but they are still weak against other types of attacks such as geometric distortions because of their high computational complexity, implementation difficulties, and poor performance.

There are two major classes of detection in a watermarking system: blind detection, which does not need the original signal for detecting the mark and non-blind detection that uses the original signal in the detection process.

In [15] a new non-blind watermarking scheme is proposed for audio signals. In this paper we extend this method to two wavelet-based image watermarking scheme. In these methods we segment the original image to smaller blocks, apply wavelet transform to these blocks and embed the binary watermark data in the lowpass scale of each block.

## 2. BACKGROUND

In [15] a novel method is presented for audio watermarking. In this method the host signal in transform domains is modified with respect to the binary watermark signal (0 or 1). The embedding processes used in this method can be summarized as follows:
- Windowing the host signal.
- Applying the wavelet transform to each frame.
- Embedding the watermark bit of 1 or 0 to a number of wavelet coefficients, $W(i)$, of each frame based on the following equations:

$$W'(i) = W(i).\alpha \qquad \text{For embedding 1} \qquad (1)$$
$$W'(i) = W(i)/\alpha \qquad \text{For embedding 0} \qquad (2)$$

where $\alpha$ is he strength factor.
- Inverse wavelet transforming the resulted coefficients of each frame.

In the detection process, the embedded data is detected by the following process:
- Windowing the original and received signal and Applying the wavelet transform to each frame of them.
- Calculating a compare vector by dividing the wavelet coefficients of the received signal to the original one.
- Detecting the embedded bit by comparing the vector which is calculated in the second step with a threshold level. If the majority of the vector components is larger than the threshold level, the embedded bit would be 1, otherwise 0 is detected

It is proved in [15] that as the embedding process is symmetrical, the best threshold value is $\frac{1}{2}\left(\alpha + \frac{1}{\alpha}\right)$.

## 3. PROPOSED METHODS

Now we describe our proposed image watermarking scheme. We extend the audio watermarking method shown in [15] to two image watermarking techniques.

### 3.1 Method 1

In the first approach we propose a simple watermarking scheme.

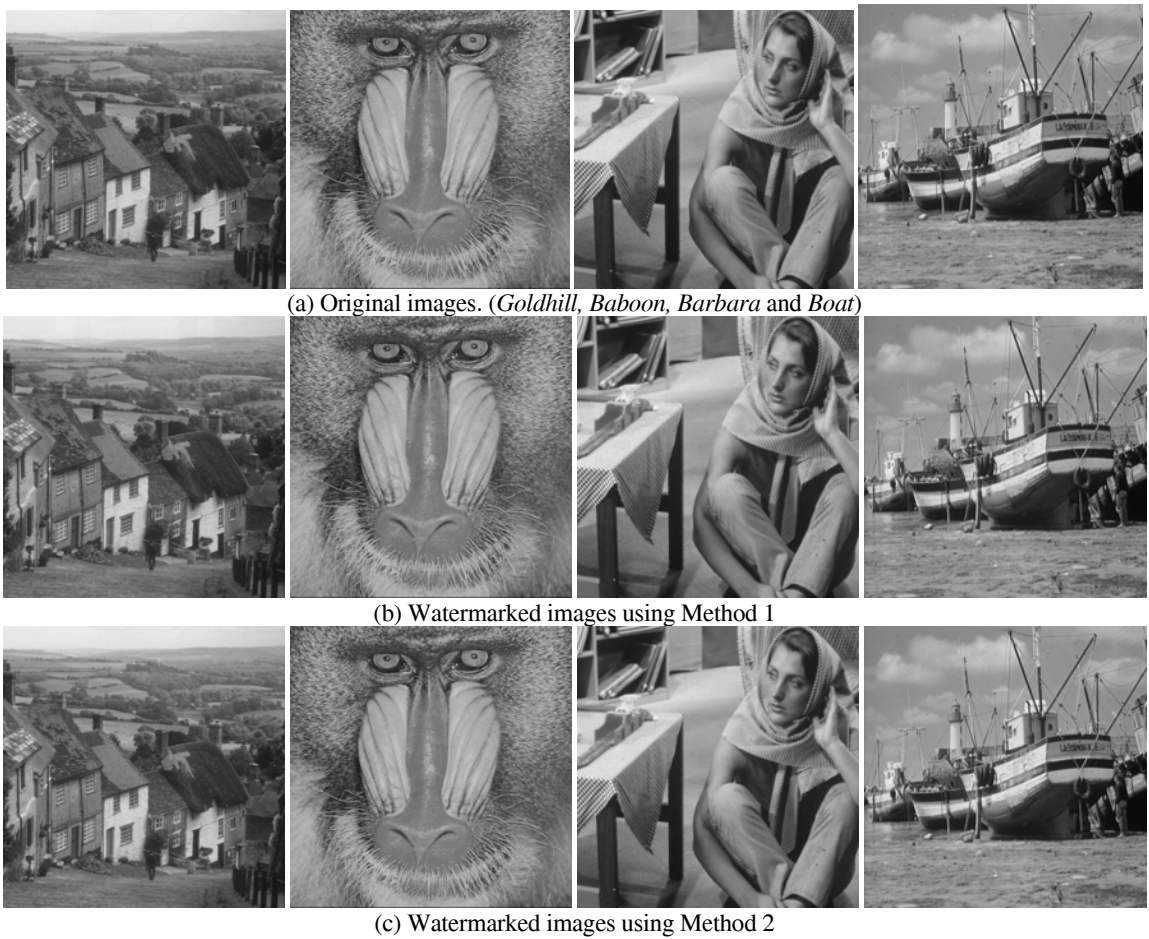

(a) Original images. (*Goldhill, Baboon, Barbara* and *Boat*)

(b) Watermarked images using Method 1

(c) Watermarked images using Method 2

Figure 1 – Original and Watermarked images.

### 3.1.1 Watermark embedding

The embedding processes used in this method can be summarized as:
- Segmenting the original image to small non-overlapping blocks and then applying the wavelet transforming to each block.
- In each block, the wavelet coefficients in the last lowpass scale are modified for embedding 1 or 0 based on (1) and (2).
- Applying the inverse wavelet transform to the obtained wavelet coefficients.

### 3.1.2 Watermark detection

The detection process can be described as follows:
- Block segmenting of the received and the original image and applying the wavelet transform to each block for the both images.
- Calculating the comparing matrix by dividing the wavelet coefficients of the received image to the original one.
- Detecting the embedded bit by comparing the matrix which is calculated in the second step with a threshold level. If the majority of the matrix components is larger than the threshold level, the embedded bit would be 1, otherwise 0 is detected.

We use the threshold level of $\frac{1}{2}\left(\alpha+\frac{1}{\alpha}\right)$, same as [15].

### 3.2 Method 2

In the second approach we propose a modified version of the first method.

### 3.2.1 Watermark embedding

The embedding process in this method is the same as method one except that we perform an additional task after block segmentation of the original image:
- The variance of each block is calculated and the first $N$ blocks with higher variances are selected for watermark embedding.

Then the same tasks are performed to embed watermark data in these selected blocks.

### 3.2.2 Watermark detection

In the detection step we first select $N$ blocks of the original image with higher variances and the same blocks in the received image. Then we perform the detection process same as before on these selected blocks. Using this approach we can choose a larger strength factor α, which increases the robustness while maintains the imperceptibility.

## 4. EXPERIMENTAL RESULTS

In this section we perform several experiments to test the proposed algorithms and evaluated its performance against various kinds of attacks. Throughout our experiments we choose the strength factor of $\alpha = 1.01$ for Method 1 and $\alpha = 1.025$ for Method 2. These factors are selected to maximize the robustness of this approach, while the modifications introduced by the watermarking process are imperceptible. Also we use the Daubechies length-8 symlet filters with maximum levels of decomposition to compute the 2-D DWT. It means that for $N \times N$ block size we use $log_2 N$ levels of decomposition. All the results are obtained by averaging on five runs. We also use a pseudorandom binary sequence as the watermarking signal. Throughout the experiments we mention Method 1 as M1 and Method 2 as M2.

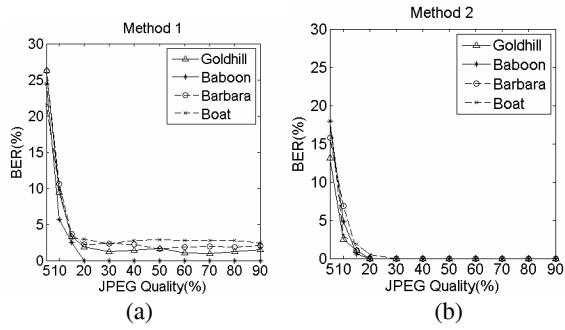

Figure 2 – BER(%) results after JPEG attack with various qualities.

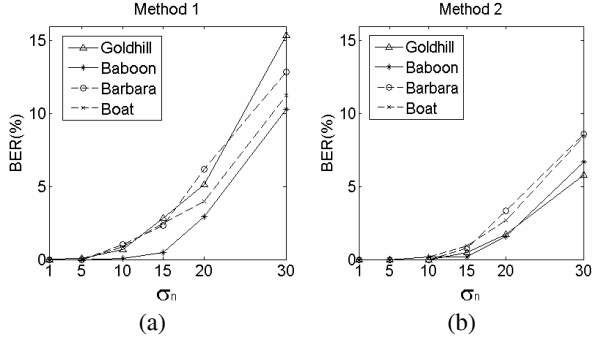

Figure 3 – BER(%) results after Noise attack for different noise standard deviations $\sigma_n$

Table 1. BER(%) resulted for different rotation angles.

| Image | Method | 0.5° | -.05° | 1° | -1° | 5° | -5° | 10° | 30° |
|---|---|---|---|---|---|---|---|---|---|
| Goldhill | M1 | 0.78 | 0.78 | 5.70 | 6.33 | 6.95 | 6.64 | 8.59 | 9.77 |
| | M2 | 0 | 0 | 0.31 | 1.56 | 0.94 | 1.56 | 0.94 | 2.19 |
| Baboon | M1 | 5.16 | 5.31 | 4.53 | 5.00 | 5.86 | 6.25 | 7.62 | 11.91 |
| | M2 | 0 | 0 | 3.75 | 5.63 | 5.94 | 7.81 | 9.69 | 12.03 |
| Barbara | M1 | 5.23 | 5.47 | 5.47 | 5.70 | 6.25 | 6.45 | 8.98 | 12.89 |
| | M2 | 0 | 0 | 3.59 | 3.13 | 5.78 | 5.00 | 5.00 | 13.13 |
| Boat | M1 | 5.39 | 5.55 | 5.23 | 5.78 | 5.86 | 6.05 | 9.96 | 12.30 |
| | M2 | 0 | 0 | 0 | 1.56 | 0.47 | 2.5 | 0.16 | 0 |

Table 2. BER(%) resulted for different scaling factors (SF).

| Image | Method | SF=0.9 | SF=0.8 | SF=0.7 | SF=0.6 | SF=0.5 |
|---|---|---|---|---|---|---|
| Goldhill | M1 | 11.72 | 9.18 | 7.03 | 7.91 | 8.98 |
| | M2 | 15.47 | 18.59 | 12.19 | 18.59 | 13.44 |
| Baboon | M1 | 6.25 | 9.38 | 12.03 | 13.48 | 10.74 |
| | M2 | 16.41 | 15.63 | 14.69 | 19.22 | 16.72 |
| Barbara | M1 | 14.06 | 14.84 | 11.91 | 7.62 | 14.45 |
| | M2 | 18.44 | 20.16 | 19.38 | 17.97 | 25.16 |
| Boat | M1 | 10.55 | 10.74 | 6.45 | 8.79 | 12.11 |
| | M2 | 27.34 | 30.78 | 18.12 | 20.16 | 25.62 |

Table 3. BER(%) resulted for mean and median filtering attacks with different window sizes.

| Image | Method | Mean Filter | | | Median Filter | | |
|---|---|---|---|---|---|---|---|
| | | 3×3 | 5×5 | 7×7 | 3×3 | 5×5 | 7×7 |
| Goldhill | M1 | 7.42 | 12.30 | 14.45 | 2.93 | 12.11 | 18.95 |
| | M2 | 0 | 2.61 | 6.51 | 0.52 | 5.21 | 17.19 |
| Baboon | M1 | 8.20 | 11.33 | 14.65 | 13.48 | 20.51 | 26.17 |
| | M2 | 2.61 | 9.11 | 14.06 | 4.69 | 19.01 | 22.14 |
| Barbara | M1 | 11.91 | 15.82 | 15.82 | 1.95 | 11.91 | 15.23 |
| | M2 | 5.73 | 8.33 | 11.46 | 3.91 | 14.84 | 23.44 |
| Boat | M1 | 6.45 | 12.11 | 15.43 | 3.52 | 10.35 | 19.73 |
| | M2 | 0.78 | 2.61 | 11.46 | 3.91 | 16.41 | 17.44 |

We use BER (bit-error-rate) to measure the performance of the watermarking scheme against attacks. A set of four common images were tested for our experiments. The images are illustrated in Figure 1.a. They are all 512×512 standard images including: *Goldhill, Baboon, Barbara* and *Boat*. Their watermarked versions using Method 1 and Method 2 are shown in Figures 1.b and 1.c. As we see the imperceptibility of the watermarked images are satisfied for both methods. The mean PSNR (peak-signal-to-noise-ratio) of the watermarked images are 46.45dB, 45.93dB, 46.26dB and 45.60dB respectively for method1 and 47.34dB, 47.65dB, 48.04dB and 47.79dB respectively for method2.

In the second experiment, we test the robustness of the proposed watermarking methods against JPEG compression attacks. Figure 2 shows the resulted BER for different images. As we see, both of our proposed methods, especially Method 2 are highly robust against JPEG attacks.

In the third experiment, the watermarks are tested against additive white Gaussian noise attack with different noise levels. The BER results are shown in Figure 3. Again, we see that the proposed watermarking schemes are robust against even high variance noise attack.

Robustness against rotation is the concept of the fourth experiment. Using template matching the rotation attack can be compensated by identifying the rotation angle and then rotating the image back [8, 16 and 17]. Therefore, the introduced distortion only comes from the interpolation due to image rotation. The BER results are shown in Table 1. As we see even for large rotation angles our watermarking methods are still robust.

In the next experiment we test scaling attack. In this case we should first restore the image to its original size and then perform the watermark detection process. Most of the watermarking schemes are less robust against this attack. The results of our watermarking schemes for scaling attack with different scaling factors are shown in Table 2. As we see this attack could cause much distortion in watermark signal but our method is still robust and could detect the watermark successfully. Also we can see that in this case the Method 1 has a better performance than the Method 2, which is due to the larger block size in Method 1.

We investigate the robustness of the proposed methods against filtering attack in the sixth experiment. In this experiment mean filter as a linear filter and median filter as a nonlinear filter with different window sizes are tested. The results can be seen in Table 3. We see that although these filters highly distort the image and degrade its quality, but our watermarking techniques, especially the second one, are still robust against these attacks and could detect the watermark signal successfully.

As the last part of this section, we compare the proposed method with several established image watermarking techniques. These methods include: Dugad's wavelet based spread spectrum method [9], the improved SS (ISS) method [10], the holographic method [11], EPCM [8], the multistage VQ (MVQ) method [12], P&Z method [13] and K&R method [14]. We test these algorithms against JPEG attack using *Lena* image. In some of these papers correlation coefficient is used to measure the watermark robustness against attacks. Correlation coefficient is defines as:

$$\rho(w, w') = \frac{\sum w(n)w'(n)}{\sqrt{\sum w(n)^2 \sum w'(n)^2}} \quad (3)$$

where $w$ is the watermark signal which is a string of {+1 and -1} and $w'$ is the extracted signal. Table 5 shows the result of these methods along with our proposed methods. As we see both of the proposed methods outperforms all of these established techniques. Also we see that the Method 2 is more robust than Method 1.

Table 4. Robustness comparison of several methods against JPEG compression attack for *Lena* image. (CC: Correlation coefficient)

| Method | Measure | JPEG Quality (%) | | | | | | | | |
|---|---|---|---|---|---|---|---|---|---|---|
| | | 10 | 20 | 30 | 40 | 50 | 60 | 70 | 80 | 90 |
| Dugads | CC | 0.78 | .86 | 1 | 1 | 1 | 1 | 1 | 1 | 1 |
| P&Z | CC | 0.1 | .18 | .25 | .34 | .42 | .49 | .64 | .80 | .94 |
| KR | CC | 0.05 | .16 | .18 | .26 | .32 | .45 | .54 | .68 | .88 |
| M1 | CC | **0.82** | **.99** | **.99** | **1** | **1** | **1** | **1** | **1** | **1** |
| M2 | CC | **0.92** | **1** | **1** | **1** | **1** | **1** | **1** | **1** | **1** |
| ISS | BER(%) | 8 | 1.8 | 1.8 | 0 | 0 | 0 | 0 | 0 | 0 |
| holo | BER(%) | 14 | 14 | 14 | 9.7 | 7.9 | 6.1 | 6.1 | - | - |
| ECPM | BER(%) | 1.8 | 1.8 | 0 | 0 | 0 | 0 | 0 | 0 | 0 |
| MVQ | BER(%) | - | - | 6.30 | - | 3.20 | - | - | 1.20 | - |
| M1 | BER(%) | **9.29** | **0.39** | **0.16** | **0** | **0** | **0** | **0** | **0** | **0** |
| M2 | BER(%) | **3.91** | **0** | **0** | **0** | **0** | **0** | **0** | **0** | **0** |

As the last part of this section we should mention the low computational complexity of the proposed methods. The CPU time for watermarking a 512×512 image using a 3 GHz Pentium IV machine is about 6.17 sec and 2.66 sec for Method 1 and Method 2, respectively. Also the CPU time for watermark detection of the same image is about 3.92 sec and 1.49 sec, respectively.

## 5. CONCLUSION

In this paper we presented a new wavelet-based image watermarking technique which is suitable for image copyright protection. In this method the host image was segmented to small blocks and the watermark data was embedded in the lowpass wavelet coefficients of each block with one of two methods mentioned in Section 3. The simulation results on several images confirm the imperceptibility of the watermarked image. We also showed the robustness of the proposed method against several attacks. Experiments demonstrate that Method 2 outperforms Method 1 against most of attacks. Due to the low computational complexity of the proposed algorithm it can be implemented in real time purposes. Future work may be done by trying to develop a blind watermarking scheme considering the idea of the proposed method. Moreover, we want to improve the idea of Method 2 for selecting high variance blocks for watermark embedding.